\documentclass[aps,prl,twocolumn,showpacs,superscriptaddress]{revtex4}

\usepackage{graphicx}

\begin{document}

\title{Low-frequency incommensurate magnetic response
in strongly correlated systems}

\author{A.~Sherman}

\affiliation{Institute of Physics, University of Tartu, Riia 142, 51014
Tartu, Estonia}

\author{M.~Schreiber}

\affiliation{Institut f\"ur Physik, Technische Universit\"at, D-09107
Chemnitz, Federal Republic of Germany}

\date{\today}

\begin{abstract}
It is shown that in the $t$-$J$ model of Cu-O planes at low frequencies
the dynamic spin structure factor is peaked at incommensurate wave
vectors $(\frac{1}{2}\pm\delta,\frac{1}{2})$,
$(\frac{1}{2},\frac{1}{2}\pm\delta)$. The incommensurability is
connected with the momentum dependencies of the magnon frequency and
damping near the antiferromagnetic wave vector. The behavior of the
incommensurate peaks is similar to that observed in
La$_{2-x}$(Ba,Sr)$_x$CuO$_{4+y}$ and YBa$_2$Cu$_3$O$_{7-y}$: for hole
concentrations $0.02<x\alt 0.12$ we find that $\delta$ is nearly
proportional to $x$, while for $x>0.12$ it tends to saturation. The
incommensurability disappears with increasing temperature. Generally
the incommensurate magnetic response is not accompanied by an
inhomogeneity of the carrier density.
\end{abstract}


\maketitle

One of the most interesting features of the inelastic neutron
scattering in lanthanum cuprates is that for hole concentrations $x
\alt 0.04$ the spin fluctuation scattering remains commensurate: it is
peaked at the antiferromagnetic wave vector ${\bf
Q}=(\frac{1}{2},\frac{1}{2})$ in the reciprocal lattice units $2\pi/a$
with the lattice period $a$. However, for larger $x$ and low energy
transfers the spin fluctuations become incommensurate with scattering
peaks shifted to the four symmetric positions
$(\frac{1}{2}\pm\delta,\frac{1}{2})$ and
$(\frac{1}{2},\frac{1}{2}\pm\delta)$ \cite{Yoshizawa}. For $x\alt 0.12$
the incommensurability parameter $\delta$ is approximately equal to $x$
and varies nearly linearly with the temperature of the superconducting
transition $T_c$ \cite{Yamada}. For larger $x$ the parameter saturates
near the value $\delta\approx 0.12$. The incommensurate response was
observed both below and above $T_c$ \cite{Mason93}. There are no clear
experimental indications that in La$_{2-x}$Sr$_x$CuO$_4$ the {\em
dynamic} magnetic incommensurability is connected with lattice
distortions. However, for $x\approx 0.12$ when the {\em static}
incommensurability is observed for temperatures $T\alt T_c$ a
remarkable softening of longitudinal sound waves and soft optical
phonons were observed \cite{Suzuki}. This softening is a manifestation
of the incipient structural transition from the low-temperature
orthorhombic to the low-temperature tetragonal phase and the appearance
of the static incommensurability is connected with the pinning of the
dynamic incommensurate spin correlations by this lattice instability
\cite{Suzuki}.

The experimental data suggest that the magnetic incommensurability
emerges due to doped holes. Several approaches have been used to
clarify the underlying mechanism. As possible candidates the Fermi
surface nesting \cite{Bulut}, mean-field spiral \cite{Normand} and
stripe \cite{Hizhnyakov} phases, and phenomenological marginal Fermi
liquid \cite{Littlewood} were considered. In works based on the Fermi
surface nesting the difference in the low-frequency magnetic response
in La$_{2-x}$Sr$_x$CuO$_4$ and YBa$_2$Cu$_3$O$_{7-y}$ was attributed to
the difference in their Fermi surfaces calculated in the mean-field
approximation. However, recently the incommensurability which is
analogous to that observed in lanthanum cuprates was also detected in
YBa$_2$Cu$_3$O$_{7-y}$ \cite{Arai}. This result indicates that the
low-frequency incommensurate magnetic response may be a common property
of cuprates. This is especially interesting because the overall
magnetic response of these two classes of crystals differs essentially.
As known, in YBa$_2$Cu$_3$O$_{7-y}$ and some other high $T_c$ cuprates
the resonance peak is observed \cite{Fong} at somewhat larger
frequencies $\omega=20-40$~meV than the frequencies of incommensurate
response, while no such peak was detected in lanthanum cuprates
\cite{Mason96}. In works based on the stripe mechanism the magnetic
incommensurability is connected with the appearance of the charge
density wave. In neutron scattering such a wave is observed only in the
low-temperature tetragonal or the low-temperature less-orthorhombic
phases where the wave is stabilized by the corrugated pattern of the
in-plane lattice potential \cite{Fujita}. Among lanthanum cuprates
La$_{2-x}$Ba$_x$CuO$_4$ and La$_{2-y-x}$Nd$_y$Sr$_x$CuO$_4$ have these
phases. In La$_{2-x}$Sr$_x$CuO$_4$ in the low-temperature orthorhombic
phase no indications of the charge density wave were observed in
neutron scattering \cite{Kimura}.

In this letter we demonstrate that the low-frequency incommensurate
response can be interpreted based on the $t$-$J$ model of Cu-O planes.
We use the general form of the spin spectral function $B({\bf
k}\omega)$ obtained \cite{ShermanPRB} in Mori's projection operator
technique. This approach allows us to take proper account of strong
electron correlations inherent in cuprates. In calculating the magnon
damping the known shape of the spin-polaron band \cite{Izyumov} which
forms the Fermi surface at low $x$ was used. The mechanism of the
appearance of the incommensurability is the following: at low
frequencies $B({\bf k}\omega)$ is reduced to the quotient of the magnon
damping $\Gamma({\bf k})$ and the fourth power of the magnon frequency
$\omega_{\bf k}$. Near {\bf Q} this frequency is an increasing function
of $|{\bf k-Q}|$. Thus, $B({\bf k}\omega)$ will be peaked at
incommensurate momenta if $\Gamma({\bf k})$ is also an increasing
function of $|{\bf k-Q}|$. This takes place indeed because the magnon
damping connected with the decay into an electron-hole pair is
approximately proportional to ${\cal S}=\sum_{\bf
k'}[1-n_F(\varepsilon_{\bf k+k'}-\mu)]n_F(\varepsilon_{\bf k'}-\mu)$
where $n_F(\omega)=[\exp(\omega/T)+1]^{-1}$, $\varepsilon_{\bf k}$ is
the hole dispersion and $\mu$ the chemical potential. Due to the
approximate periodicity of $\varepsilon_{\bf k}$ with the period {\bf
Q} (connected with the short-range antiferromagnetic order) ${\cal
S}=0$ for ${\bf k=Q}$ and grows with growing $|{\bf k-Q}|$.

The spin spectral function, connected with the retarded spin Green's
function $D({\bf k}\omega)=\langle\langle s^z_{\bf k}|s^z_{\bf
-k}\rangle\rangle$ by the relation $B({\bf k}\omega)=-\pi^{-1}{\rm
Im}\,D({\bf k}\omega)$, reads \cite{ShermanPRB}
\begin{equation}\label{ssf}
B({\bf k}\omega)=-\frac{\pi^{-1}\omega\Gamma({\bf
k}\omega)}{[\omega^2-f^{-1}\omega\Theta({\bf k}\omega)-\omega^2_{\bf
k}]^2+f^{-2}\omega^2\Gamma^2({\bf k}\omega)},
\end{equation}
where $s^z_{\bf k}$ is the $z$-component of the spin-$\frac{1}{2}$
operator describing localized Cu spins, $f=4JC_1(\gamma_{\bf k}-1)$
with $\gamma_{\bf k}=\frac{1}{4}\sum_{\bf a}\exp(i{\bf ka})$, {\bf a}
are four vectors connecting nearest neighbor sites in the square
lattice of the 2D $t$-$J$ model, $J$ is its exchange parameter,
$C_1=2\langle s^z_{\bf n}s^z_{\bf n+a}\rangle$ is the spin correlation
on the neighbor sites {\bf n} and {\bf n+a} with the statistical
averaging denoted by the angular brackets. In the normal state the
magnon frequency and the damping are given by the relations
\begin{equation}\label{frequency}
\omega^2_{\bf k}=16\alpha J^2|C_1|(1-\gamma_{\bf
k})(\Delta+1+\gamma_{\bf k}),
\end{equation}
\begin{eqnarray}
\Gamma({\bf k}\omega)&=&\frac{32\pi t^2J^2}{N^2}
\frac{1-\exp(\omega/T)}{\omega}\nonumber\\
&\times&\sum_{{\bf k}_1{\bf k}_2}g^2_{{\bf kk}_1{\bf k}_2}
\int\!\!\!\!\int^\infty_{-\infty}d\omega_1d\omega_2n_B(\omega_2)
\nonumber\\
&\times&[1-n_F(\omega_1)]n_F(\omega+\omega_1-\omega_2)B({\bf
k}_2\omega_2)\nonumber\\
&\times&A({\bf k}_1\omega_1)A({\bf k+k}_1-{\bf
k}_2,\omega+\omega_1-\omega_2), \label{damping}
\end{eqnarray}
where $t$ is the hopping parameter of the $t$-$J$ model, $N$ is the
number of sites, $\alpha$ is the vertex parameter \cite{ShermanEPJB}
which is of the order of unity, $g_{{\bf kk}_1{\bf k}_2}=(\gamma_{{\bf
k}_2}+\frac{1}{4})(\gamma_{{\bf k}_2-{\bf k}_1}-\gamma_{{\bf
k}_1}-\gamma_{{\bf k+k}_1-{\bf k}_2}+\gamma_{{\bf k+ k}_1})$,
$n_B(\omega)=[\exp(\omega/T)-1]^{-1}$, and $A({\bf
k}\omega)=-\pi^{-1}{\rm Im}\,\langle\langle a_{\bf
k\sigma}|a^\dagger_{\bf k\sigma}\rangle\rangle$ is the hole spectral
function with the operator $a^\dagger_{\bf k\sigma}$ creating a hole
with the spin projection $\sigma=\pm 1$. In Eq.~(\ref{frequency}), the
parameter $\Delta$ describes the spin gap at {\bf Q} (see
\cite{ShermanEPJB} and references therein). The correction $\Theta({\bf
k}\omega)$ to the magnon frequency  can be obtained from $\Gamma({\bf
k}\omega)$ and the Kramers-Kronig relation. The dynamic structure
factor measured in neutron scattering experiments is connected with
$B({\bf k}\omega)$ by the relation $S({\bf
k}\omega)=\pi[1-\exp(-\omega/T)]^{-1}B({\bf k}\omega)$ \cite{Kastner}.

If the magnons near {\bf Q} are not overdamped, the dominant feature of
the spectral function (\ref{ssf}) is the maximum near
$\omega=\omega_{\bf Q}=4J(2\alpha|C_1|\Delta)^{1/2}$. This maximum, the
resonance peak, is observed in YBa$_2$Cu$_3$O$_{7-y}$ and some other
high $T_c$ cuprates \cite{Fong}. Notice that its frequency
$\omega=20-40$~meV is close to the frequency of the spin gap
$\omega_{\bf Q}$ \cite{Morr,ShermanPRB}. In lanthanum cuprates the
magnons near {\bf Q} are apparently overdamped even in the underdoped
region and due to the small superconducting gap the resonance peak was
not observed in the superconducting state either \cite{Mason96,Morr}.
In this case the frequency dependence of the spectral function is
determined by $\Gamma({\bf k}\omega)$ rather than the resonance
denominator in Eq.~(\ref{ssf}). In lanthanum cuprates the
incommensurate response is observed for frequencies of the order of
several millielectronvolts. For such frequencies Eq.~(\ref{ssf}) can be
essentially simplified,
\begin{equation}\label{ssfsimp}
B({\bf k}\omega)=-\frac{\pi^{-1}\omega\Gamma({\bf
k}\omega)}{\omega^4_{\bf k}}.
\end{equation}
As seen from Eq.~(\ref{frequency}), $\omega_{\bf k} \approx
[\omega^2_{\bf Q}+c^2({\bf k-Q})^2]^{1/2}$ near {\bf Q} and therefore
in Eq.~(\ref{ssfsimp}) $\omega_{\bf k}^{-4}$ decreases rapidly with
increasing $|{\bf k-Q}|$. In this case for the function $B({\bf
k}\omega)$ to be peaked at incommensurate momenta near {\bf Q} the
numerator of the fraction (\ref{ssfsimp}), the magnon damping, has to
be a growing function of $|{\bf k-Q}|$.
\begin{figure}
\includegraphics[width=4cm]{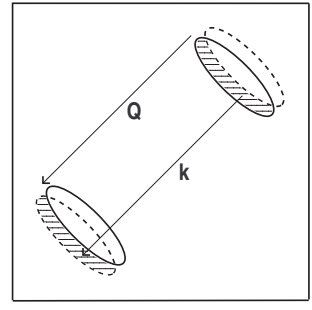}
\caption{The Brillouin zone of the square lattice. Solid curves are two
of four ellipses forming the Fermi surface of the band
(\protect\ref{spb}) at small $x$. Dashed lines are the Fermi surface
contours shifted by $\pm({\bf k-Q})$. Regions of ${\bf k}_1$ and ${\bf
k+k}_1$ contributing to the damping (\protect\ref{apprdamping}) are
shaded.}\label{Fig_i}
\end{figure}

To make sure that the damping may really have such a behavior let us
consider the case of low hole concentrations when in
Eq.~(\ref{damping}) the hole and spin spectral functions can be
approximated as \cite{ShermanEPJB}
\begin{eqnarray}
A({\bf k}\omega) &=& \phi\delta(\omega-\varepsilon_{\bf k}+\mu),
 \nonumber \\
B({\bf k}\omega) &=& \frac{1}{2}\sqrt{\frac{|C_1|}{\alpha}}
 \sqrt{\frac{1-\gamma_{\bf k}}{\Delta+1+\gamma_{\bf k}}}
 \label{AnB}\\
&&\times [\delta(\omega-\omega_{\bf k})-\delta(\omega+\omega_{\bf
 k})], \nonumber
\end{eqnarray}
where $\phi=\frac{1}{2}(1+x)$ and the dispersion of the spin-polaron
band \cite{Izyumov,ShermanEPJB}
\begin{eqnarray}
\varepsilon_{\bf k}/t &=& -2.3893+0.055358[\cos(2k_x)+\cos(2k_y)]
 \nonumber\\
&&+0.17857\cos(k_x)\cos(k_y). \label{spb}
\end{eqnarray}
The approximate formula (\ref{spb}) reproduces the main known features
of the spin-polaron band \cite{Izyumov}. In particular, it has minima
at $(\pm\pi/2,\pm\pi/2)$. The bottom of the band, $-2.5t$, and its
width, $0.4t=2J$, are set to be close to the values obtained in
spin-wave and exact diagonalization calculations for $J=0.2t$. This
parameter ratio is close to that observed in hole-doped cuprates
\cite{McMahan}. Also the shape of the band (\ref{spb}) and its Fermi
surface are similar to those observed in oxychlorides and in
La$_{2-x}$Sr$_x$CuO$_4$ at light and moderate doping
\cite{Damascelli,ShermanEPJB}. For low $x$ the Fermi surface of the
band (\ref{spb}) consists of four ellipses centered at
$(\pm\pi/2,\pm\pi/2)$. Two of them are shown in Fig.~\ref{Fig_i}.
Substituting Eq.~(\ref{AnB}) in Eq.~(\ref{damping}) and taking into
account that for ${\bf k \approx Q}$ and small $\omega$ the region
around {\bf Q} makes the main contribution in the summation over ${\bf
k}_2$ we find
\begin{eqnarray}
\Gamma({\bf k}\omega) &\approx& \frac{32\pi t^2J^2\phi^2}{N^2}
 \frac{1-\exp(\omega/T)}{\omega}\nonumber\\
&\times& \sum_{{\bf k}_1}g^2_{{\bf kk}_1{\bf
 Q}}[1-n_F(\varepsilon_{{\bf
 k}_1}-\mu)]n_F(\varepsilon_{{\bf k+k}_1}-\mu)\nonumber\\
&\times& \sum_{{\bf k}_2}B({\bf k}_2,\omega+\varepsilon_{{\bf k}_1}
 -\varepsilon_{{\bf k+k}_1-{\bf k}_2})\nonumber\\
&\times& n_B(\omega+\varepsilon_{{\bf
 k}_1}-\varepsilon_{{\bf k+k}_1-{\bf k}_2}). \label{apprdamping}
\end{eqnarray}
The dependence $\Gamma({\bf k})$ is mainly determined by the sum ${\cal
S}=\sum_{{\bf k}_1}g^2_{{\bf kk}_1{\bf Q}}[1-n_F(\varepsilon_{{\bf
k}_1}-\mu)]n_F(\varepsilon_{{\bf k+k}_1}-\mu)$. For low temperatures
the regions of ${\bf k}_1$ and ${\bf k+k}_1$ contributing to the sum
are shaded in Fig.~\ref{Fig_i}. The damping is roughly proportional to
the area of these regions. Thus, for low $T$ and $x$ the damping grows
with growing $|{\bf k-Q}|$ which shifts the peaks in $B({\bf k}\omega)$
to incommensurate positions.
\begin{figure}
\includegraphics[width=6cm]{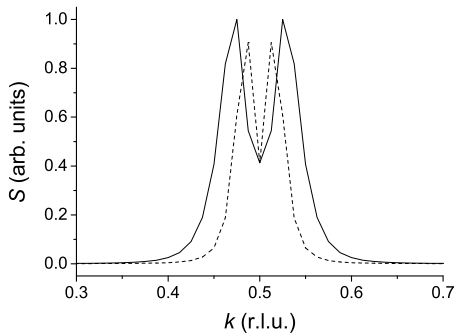}
\caption{The dynamic structure factor vs.\ wave vector along the edge
(solid line) and the diagonal (dashed line) of the Brillouin zone for
$x=0.06$, $T=0.01t$, and $\omega=0.004t$.}\label{Fig_ii}
\end{figure}

Key elements for the validity of the above conclusion are the spin
spectral function in Eq.~(\ref{damping}) which is strongly peaked near
{\bf Q} and an approximate periodicity of the hole dispersion with the
period {\bf Q}. Due to the doubling of the elementary cell in
antiferromagnetically ordered crystals both these conditions have to be
fulfilled in lightly and moderately doped cuprates which are
characterized by the short-range antiferromagnetic ordering. A small
damping of hole states is also necessary (the incommensurability was
suppressed in the self-consistent calculations \cite{ShermanEPJB} due
to a sizeable hole damping introduced for stabilizing the iteration
procedure).

We used Eqs.~(\ref{damping})--(\ref{spb}) and parameters $\alpha$,
$C_1$, and $\Delta$ obtained self-consistently \cite{ShermanEPJB} for
calculating the dynamic structure factor $S({\bf k}\omega)$. We
considered lattices up to 120$\times$120 sites with the parameters
$J=0.2t$. Figure~\ref{Fig_ii} demonstrates the dynamic structure factor
calculated along edge and diagonal of the Brillouin zone near {\bf Q}.
The parameters chosen are close to those used in experiments: for
$t=0.5$~eV, $T \approx 58$~K and $\omega=2$~meV. In agreement with
experimental observations \cite{Yoshizawa,Yamada,Mason93} peaks along
the edge are more intensive than those along the diagonal.

\begin{figure}
\includegraphics[width=7cm]{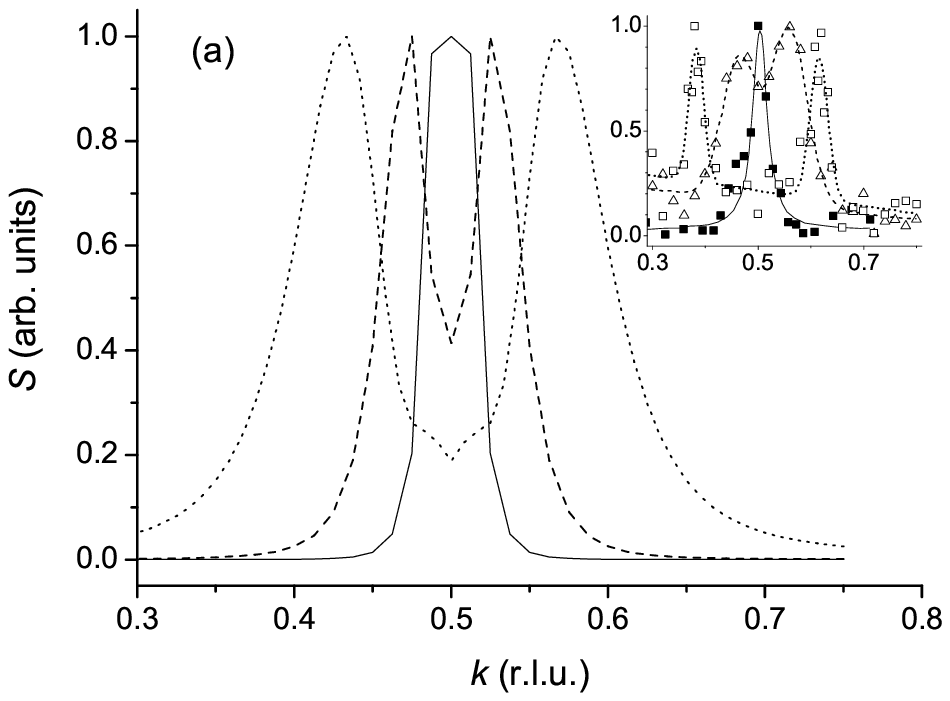}

\vspace{1ex}\includegraphics[width=7cm]{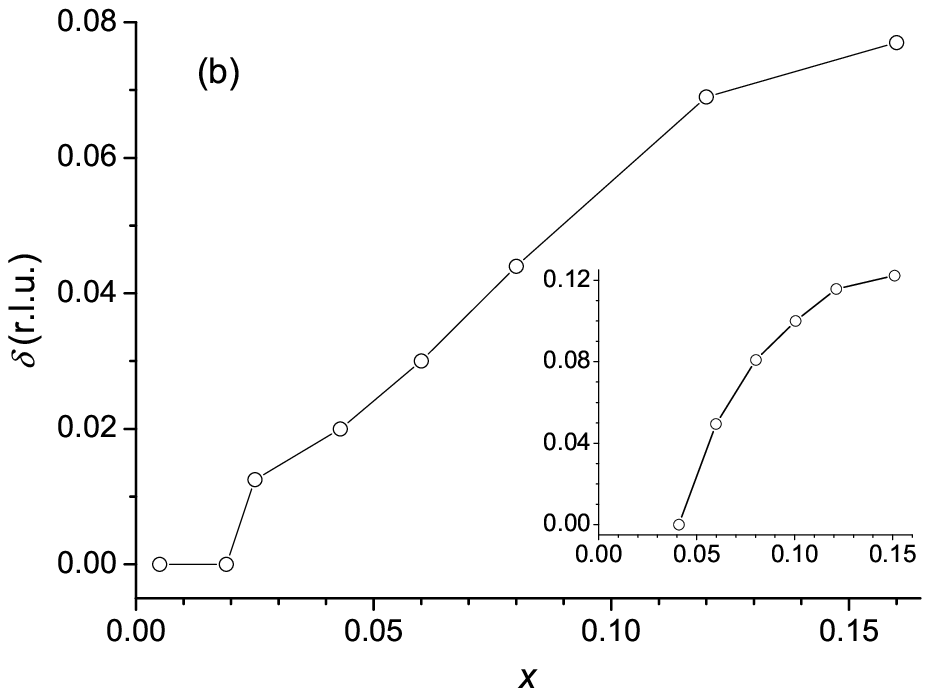} \caption{(a) The
dynamic structure factor vs.\ wave vector along the edge for $x=0.015$
(solid line), 0.06 (dashed line), and 0.12 (dotted line). For all
curves $T=0.01t$ and $\omega=0.004t$. Inset: the structure factor
measured \protect\cite{Yamada,Wakimoto} in La$_{2-x}$Sr$_x$CuO$_4$ for
$x=0.03$, $T=2$~K, $\omega=0$ (filled squares and solid line),
$x=0.06$, $T=25$~K, $\omega=2$~meV (triangles and dashed line), and
$x=0.12$, $T=31$~K, $\omega=2$~meV (open squares and dotted line).
Fitted curves in the inset are from \protect\cite{Yamada,Wakimoto}. (b)
The incommensurability parameter $\delta$ vs.\ $x$ for $T=0.01t$ and
$\omega=0.004t$. Inset: experimental data \protect\cite{Yamada} for
La$_{2-x}$Sr$_x$CuO$_4$. Connecting lines are a guide to the
eye.}\label{Fig_iii}
\end{figure}

Figure~\ref{Fig_iii} shows the concentration dependence of the peak
position along the edge.
Starting from $x \approx 0.02$ there appears the incommensurability --
instead of one peak at {\bf Q} four peaks are observed at
$(\frac{1}{2},\frac{1}{2}\pm\delta)$ and
$(\frac{1}{2}\pm\delta,\frac{1}{2})$. For $x\alt 0.12$ the
incommensurability parameter $\delta$ grows nearly linearly with
growing $x$ and tends to saturation for $x>0.12$. As seen from the
insets in Fig.~\ref{Fig_iii}, this behavior is similar to the
experimental observations except that in experiment the
incommensurability arises at $x \approx 0.04$ and the parameter
$\delta$ is approximately 1.5 times larger. In our calculations its
concentration dependence is mainly determined by the variation of the
spin-gap parameter $\Delta$ in $\omega_{\bf k}$ in Eq.~(\ref{ssfsimp}).
For $x<0.12$ this parameter grows with $x$ and saturates at $x \approx
0.12$ \cite{ShermanEPJB}.

\begin{figure}
\includegraphics[width=7.8cm]{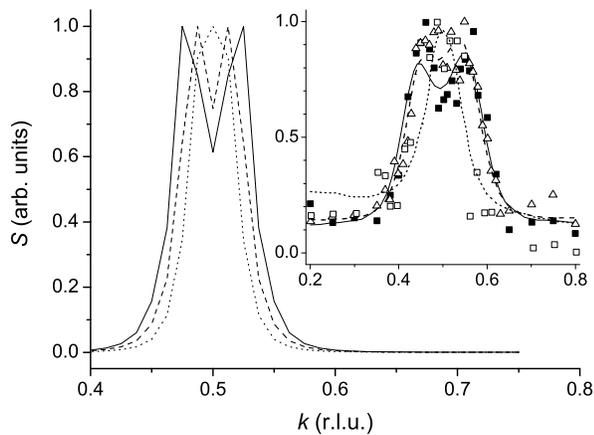}
\caption{The dynamic structure factor vs.\ wave vector along the edge
for $x=0.04$, $\omega=0.004t$ and $T=0.01t$ (solid line), $0.02t$
(dashed line), and $0.04t$ (dotted line). Inset: the structure factor
in La$_{1.93}$Sr$_{0.07}$CuO$_4$ measured \protect\cite{Hiraka} for
$\omega=4$~meV, $T=23$~K (filled squares and solid line), $T=100$~K
(triangles and dashed line), and for $\omega=8$~meV, $T=293$~K (open
squares and dotted line). Fitted curves in the inset are from
\protect\cite{Hiraka}.}\label{Fig_iv}
\end{figure}

The temperature variation of the peaks is shown in Fig.~\ref{Fig_iv}.
With growing temperature $\delta$ decreases and finally the
incommensurability disappears. As seen in the inset of
Fig.~\ref{Fig_iv}, the analogous behavior of the dynamic structure
factor is observed experimentally \cite{Arai,Hiraka}. Notice that using
Eq.~(\ref{AnB}) we neglected completely the damping in the hole
spectral function. As mentioned, this damping decreases or even
suppresses the incommensurability. The thermal growth of the damping
hastens the decrease of the incommensurability with temperature.

It should be emphasized that at least in the case when magnons near
{\bf Q} are not overdamped the low-frequency incommensurability in
$B({\bf k}\omega)$ influences only weakly the hole spectral functions.
The solution of the self-energy equations \cite{ShermanEPJB} for the
$t$-$J$ model indicates that in these conditions the density of holes
remains homogeneous as it is apparently observed in the low-temperature
orthorhombic phase in La$_{2-x}$Sr$_x$CuO$_4$ \cite{Kimura}.

In summary, it was shown that for hole concentrations $x \agt 0.02$ the
low-frequency magnetic response of the two-dimensional $t$-$J$ model is
incommensurate. The incommensurability is the consequence of the local
minimum in the momentum dependence of the magnon damping. The minimum
is located at the antiferromagnetic wave vector ${\bf
Q}=(\frac{1}{2},\frac{1}{2})$ in the reciprocal lattice units and is
connected with the growth of the phase space for the magnon damping
into the electron-hole pair with distance of the wave vector {\bf k}
from {\bf Q}. For such behavior the spin spectral function has to be
strongly peaked near {\bf Q} and the hole dispersion has to be
approximately periodic with the period {\bf Q}. Both these conditions
are fulfilled in crystals with short-range antiferromagnetic order to
which lightly and moderately doped cuprates belong. In agreement with
experiment for $x \alt 0.12$ the incommensurability grows nearly
proportional to $x$ and tends to saturation for $x>0.12$. This
dependence is connected with the concentration dependence of the spin
gap at {\bf Q}. Also in agreement with experiment the
incommensurability decreases with increasing temperature. In the
considered mechanism the magnetic incommensurability is not accompanied
by the inhomogeneity of the carrier density.

This work was partially supported by the ESF grant No.~5548 and by DFG.


\begin{thebibliography}{99}
\bibitem{Yoshizawa}H.~Yoshizawa {\it et al.}, J.\ Phys.\ Soc.\ Jpn.\
{\bf 57}, 3686 (1988); R.~J.~Birgeneau {\it et al.}, Phys.\ Rev.\ B
{\bf 39}, 2868 (1989).

\bibitem{Yamada}K.~Yamada {\it et al.}, Phys.\ Rev.\ B {\bf 57}, 6165
(1998).

\bibitem{Mason93}T.~E.~Mason {\it et al.}, Phys.\ Rev.\ Lett.\ {\bf 71},
919 (1993); M.~Matsuda {\it et al.}, Phys.\ Rev.\ B {\bf 49}, 6958
(1994).

\bibitem{Suzuki}T.~Suzuki {\it et al.}, Phys.\ Rev.\ B {\bf
57}, R3229 (1998); H.~Kimura {\it et al.}, J.\ Phys.\ Soc.\ Jpn.\ {\bf
69}, 851 (2000).

\bibitem{Bulut}N.~Bulut {\it et al.}, Phys.\ Rev.\ Lett.\ {\bf 64}, 2723
(1990); Q.~Si {\it et al.}, Phys.\ Rev.\ B {\bf 47}, 9055 (1993).

\bibitem{Normand}B.~Normand and P.~A.~Lee, Phys.\ Rev.\ B {\bf 51},
15519 (1995).

\bibitem{Hizhnyakov}V.~Hizhnyakov and E.~Sigmund, Physica C {\bf 156},
655 (1988); J.~Zaanen and O.~Gunnarson, Phys.\ Rev.\ B {\bf 40}, 7391
(1989); S.~R.~White and D.~J.~Scalapino, Phys.\ Rev.\ Lett.\ {\bf 80},
1272 (1998).

\bibitem{Littlewood}P.~B.~Littlewood {\it et al.}, Phys.\ Rev.\ B
{\bf 48}, 487 (1993).

\bibitem{Arai}M.~Arai {\it et al.}, Phys.\ Rev.\ Lett.\ {\bf 83}, 608
(1999); P.~Dai {\it et al.}, Science {\bf 284}, 1344 (1999).

\bibitem{Fong}H.~F.~Fong {\it et al.}, Phys.\ Rev.\ B {\bf 61}, 14773
(2000).

\bibitem{Mason96}T.~E.~Mason {\it et al.}, Phys.\ Rev.\ Lett.\ {\bf 77},
1604 (1996).

\bibitem{Fujita}M.~Fujita {\it et al.}, Phys.\ Rev.\ Lett.\ {\bf 88},
167008 (2002).

\bibitem{Kimura}H.~Kimura {\it et al.}, Phys.\ Rev.\ B {\bf 59},
6517 (1999).

\bibitem{ShermanPRB}A.~Sherman and M.~Schreiber, Phys.\ Rev.\ B {\bf 68},
094519 (2003).

\bibitem{Izyumov}Yu.~A.~Izyumov, Usp.\ Fiz.\ Nauk {\bf 167}, 465
(1997) [Phys.-Usp.\ (Russia) {\bf 40}, 445 (1997)]; E.~Dagotto, Rev.\
Mod.\ Phys. {\bf 66}, 763 (1994).

\bibitem{ShermanEPJB}A.~Sherman and M.~Schreiber, Eur.\ Phys.\ J.\ B
{\bf 32}, 203 (2003); Phys.\ Rev.\ B {\bf 65}, 134520 (2002).

\bibitem{Kastner}M.~A.~Kastner {\it et al.}, Rev.\ Mod.\ Phys.\ {\bf
70}, 897 (1998).

\bibitem{Morr}D.~K.~Morr and D.~Pines, Phys.\ Rev.\ Lett.\ {\bf 81},
1086 (1998).

\bibitem{McMahan}A.~K.~McMahan {\it et al.}, Phys.\ Rev.\ B {\bf 42},
6268 (1990); V.~A.~Gavrichkov {\it et al.}, Zh.\ Eksp.\ Teor.\ Fiz.\
{\bf 118}, 422 (2000) [JETP (Russia) {\bf 91}, 369 (2000)].

\bibitem{Damascelli}A.~Damascelli {\it et al.}, Rev.\ Mod.\ Phys. {\bf
75}, 473 (2003).

\bibitem{Wakimoto}S.~Wakimoto {\it et al.}, Phys.\ Rev.\ B {\bf 60},
R769 (1999).

\bibitem{Hiraka}H.~Hiraka {\it et al.}, J.\ Phys.\ Soc.\ Jpn.\ {\bf 70},
853 (2001).

\end{thebibliography}
\end{document}